\documentclass[a4paper,10pt]{scrartcl}
\usepackage[utf8]{inputenc}
\usepackage[colorlinks=true, linkcolor=red, citecolor=blue, urlcolor=blue]{hyperref}
\usepackage[backend=bibtex,style=numeric-comp,sorting=none,natbib=true,safeinputenc]{biblatex} 
\usepackage{graphicx}
\usepackage{amssymb}
\usepackage{amsmath}
\usepackage{authblk}
\usepackage{amsthm}
\usepackage{fullpage}
\usepackage{enumerate}
\usepackage[percent]{overpic}
\usepackage[page]{appendix}
\usepackage{tikz-cd}
\usepackage{marvosym}

\newcommand{\ex}[1]{{\left\langle{#1}\right\rangle}}
\newcommand{\ket}[1]{\left|{#1}\right\rangle}
\newcommand{\bra}[1]{\left\langle{#1}\right|}
\newcommand{\comment}[1]{}
\newcommand{\xddots}{%
  \raise 7pt \hbox {.}
  \mkern 3mu
  \raise 3pt \hbox {.}
  \mkern 3mu
  \raise -1pt \hbox {.}
}
\newtheorem{theorem}{Assumption}
\def\shrug{\texttt{\raisebox{0.75em}{\char`\_}\char`\\\char`\_\kern-0.5ex(\kern-0.25ex\raisebox{0.25ex}{\rotatebox{45}{\raisebox{-.75ex}"\kern-1.5ex\rotatebox{-90})}}\kern-0.5ex)\kern-0.5ex\char`\_/\raisebox{0.75em}{\char`\_}}}

\title{Epistemic Horizons: This Sentence is $\frac{1}{\sqrt{2}}\left(\ket{\mathrm{true}} + \ket{\mathrm{false}}\right)$}
\author{Jochen Szangolies\thanks{\href{mailto:jochen.szangolies@gmx.de}{\texttt{jochen.szangolies@gmx.de}}}}
\date{\vspace{-5ex}}

\begin{document}

\maketitle

\begin{abstract}

In [Found. Phys. 48.12 (2018): 1669], the notion of \emph{epistemic horizon} was introduced as an explanation for many of the puzzling features of quantum mechanics. There, it was shown that Lawvere's theorem, which forms the categorical backdrop to phenomena such as G\"odelian incompleteness, Turing undecidability, Russell's paradox and others, applied to a measurement context, yields bounds on the maximum knowledge that can be obtained about a system, which produces many paradigmatically quantum phenomena. We give a brief presentation of the framework, and then demonstrate how it naturally yields Bell inequality violations. We then study the argument due to Einstein, Podolsky, and Rosen, and show how the counterfactual inference needed to conclude the incompleteness of the quantum formalism is barred by the epistemic horizon. Similarly, the paradoxes due to Hardy and Frauchiger-Renner are discussed, and found to turn on an inconsistent combination of information from incompatible contexts.

\end{abstract}

\section{Introduction: Interpretation versus Reconstruction}\label{sec:intro}
Almost from the inception of quantum mechanics, it has been clear that it does not merely represent a theory of new phenomena, but rather, an entirely novel way of theory-building. There is now wide agreement that certain assumptions and conceptions, implicit in the Newtonian, classical framework, can no longer be upheld---albeit, and perhaps shockingly, there is as yet no consensus on what, precisely, those are.

In coming to terms with the novelty of quantum mechanics, the dominant strategy has been that of \emph{interpretation}: roughly, the attempt of matching the formalism to an underlying reality (whatever that, exactly, may mean). However, the plethora of interpretations on the market---the Wikipedia article \cite{wiki:IntQM} currently lists 14 `mainstream' interpretations---indicates that this project is still far from completion.

Sometimes, the inverse of a hard problem is more easily solved. Instead of trying to infer the underlying ontology to match the quantum formalism, one might thus take a constructive road and explore which phenomena arise naturally in certain `model' or `toy' settings, with the aim of eventually zeroing in on QM. This is the project of \emph{reconstructing} quantum mechanics: finding one or more foundational principles such that the quantum predictions naturally follow. 

In contrast to the project of interpretation, this search has, it seems, produced a significant convergence of ideas. As pointed out by Grinbaum \cite{Gr2003}, two principles are common to several recent attempts (see references in \cite{szangolies2018epistemic}):

\begin{enumerate}
 \item \label{fin}\textit{Finiteness}: There is a finite maximum of information that can be obtained about any given system. 
 \item \label{add}\textit{Extensibility}: It is always possible to acquire new information about any system.
\end{enumerate}

At first glance, these seem contradictory: how can we obtain additional information, if we already possess the maximum possible information about a system? The answer, as we will see, is closely related to one of the central puzzles of quantum mechanics: there must be a mechanism such that `old' information becomes obsolete---which, in QM, is just the hotly-debated `collapse' of the wave function.

Compare this to the situation of an observer on the spherical Earth: moving towards their horizon, bringing new terrain into view, they lose sight of what they've left behind\footnote{Although we typically expect that which has slipped beyond the horizon to remain largely unchanged, and thus, our information about it to remain accurate---but of course, this may not be the case.}.

It may nevertheless remain mysterious why nature should conspire to withhold information from us observers. To this end, in Ref.~\cite{szangolies2018epistemic}, it was proposed that the principles \ref{fin} and \ref{add} do not need to be separately postulated, but instead, follow naturally by means of applying Lawvere's fixed-point theorem \cite{Law1969} to the process of measurement, or more accurately, the prediction of measurement outcomes. 

Lawvere's theorem essentially exposes the common (categorical) structure behind phenomena such as G\"odelian incompleteness, the unsolvability of the halting problem, Russell's paradox, and many others (see \cite{Ya2003} for an overview). Thus, by connecting it to quantum measurement, unpredictability in physics---and many quantum phenomena with it---and undecidability in mathematics can be seen as two aspects of the same phenomenon: the presence of epistemic horizons.

\section{Horizons of our Understanding}\label{sec:horizons}
I do not propose to present a detailed reconstruction of the formalism of quantum mechanics here. However, I want to at least present an intuition as to how such a reconstruction, starting from the principles \ref{fin} and \ref{add}, might proceed. 

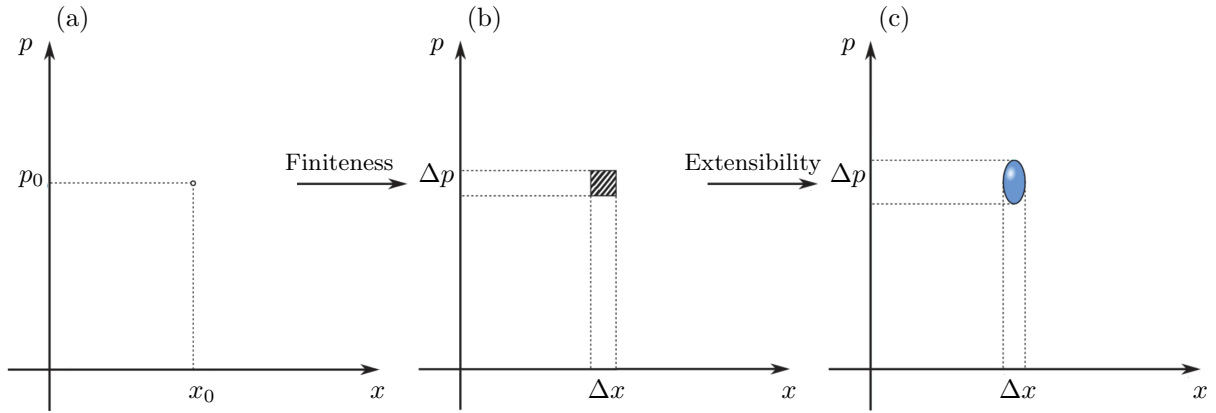
\begin{figure}[h] 
 \centering
 \begin{overpic}[width=\textwidth]{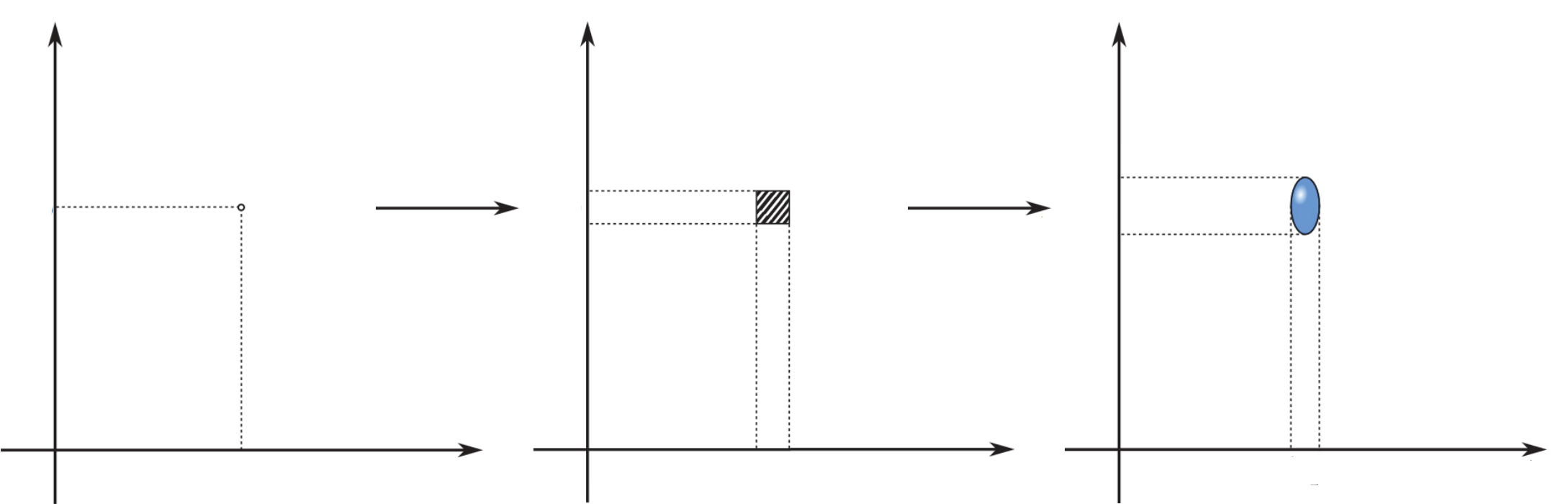}
  \put(4,32){(a)}
  \put(38,32){(b)}
  \put(72,32){(c)}
  \put(1,30){$p$}
  \put(30,1){$x$}
  \put(35,30){$p$}
  \put(64,1){$x$}
  \put(69,30){$p$}
  \put(98,1){$x$}
  \put(1,19){$p_0$}
  \put(15,1){$x_0$}
  \put(34,19){$\Delta p$}
  \put(48,1){$\Delta x$}
  \put(68,19){$\Delta p$}
  \put(82,1){$\Delta x$}
  \put(23,20){\small{Finiteness}}
  \put(56,20){\small{Extensibility}}
\end{overpic}
\caption{Quantization in phase space}
\label{pic:phase}
\end{figure}

To this end, consider as a toy model a (classical) point particle of mass $m$ moving in one dimension. Its state can be completely described by giving its position $x$ and velocity $v$---or, as is more common, its momentum $p=mv$. The space spanned by the particle's possible positions and momenta is called its \emph{phase space}. Each point in phase space gives a tuple $(x_0,p_0)$ uniquely determining the particle's state (see Fig.~\ref{pic:phase} (a)). 

From this starting point, we impose principles \ref{fin} and \ref{add}. Upon requiring that there be a maximum amount of information that can be obtained about a system, we can no longer localize its state within phase space with perfect precision---the space effectively becomes discretized (see Fig.~\ref{pic:phase} (b)). Imposing then that we can always obtain additional information entails that we can increase our information about, say, its position---but to compensate, must lose information about its momentum (see Fig.~\ref{pic:phase} (c)). 

\vspace{4mm}

Thus, we do not simply obtain a discretized phase space, but rather, there is a minimum area of localization, whose shape is determined by the information obtained about each coordinate. Since position has units of length [$\mathrm{m}$], while momentum has units of mass $\cdot$ velocity [$\mathrm{kg}\frac{\mathrm{m}}{\mathrm{s}}$], this area of maximum localizabiliy has units of [$\mathrm{kg}\frac{\mathrm{m^2}}{\mathrm{s}}$]---which is the dimension of Planck's famous constant, $\hbar$. Hence, maximum localizability in phase space is bounded by $\hbar$, which entails for the uncertainties $\Delta x$ and $\Delta p$
\begin{equation*}
    \Delta x \Delta p \gtrsim \hbar,
\end{equation*}
which is of course nothing but Heisenberg's famous uncertainty relation. In this way, assumptions \ref{fin} and \ref{add} carry us the first step of the way towards quantization.

This is, of course, an entirely heuristic picture. However, it will help, in the following, to have an intuition about the sort of project being outlined here.

\subsection{Superposition}\label{subsec:superposition}
Having now had a glimpse of how quantum phenomena emerge due to the restriction of information available about a system, it is time to consider some characteristic aspects of quantum mechanics in detail. The first step along this road will be to discuss how the impossibility of associating a definite value to every possible property of a system emerges from an argument trading on inconsistent self-reference, in much the same way as G\"odelian incompletenes \cite{Go1931} and Turing undecidability \cite{Tu1936}.

Suppose, for simplicity, that a given system $\mathcal{S}$ can be in countably\footnote{Note, however, that the argument can be generalized beyond countable sets \cite{szangolies2018epistemic}.} many different states $\{s_i\}_{i\in\mathbb{N}}$---that is, there exists an enumeration $\{s_1,s_2,\ldots\}$ of states of $\mathcal{S}$. 

Furthermore, suppose there exists likewise an enumeration of possible measurements $\{m_j\}_{j\in\mathbb{N}}$. We will suppose that these are \emph{dichotomic}: that is, each yields either $1$ or $-1$ as outcome. This is not a restriction: we can always decompose a many-valued measurement into an appropriate set of dichotomic ones. Measurements are then functions that take states as input and return values, $m_n(s_k) \in \{1,-1\}$. 

Think, as an example, of a coin: after we flip it, we make a measurement (that is, we look to see which side is up), and denote `heads' as $1$, `tails' as $-1$. For this system, there exist only two states---$s_1$ for heads and $s_2$ for tails---and one measurement $m_1$, and we have 
\begin{align*}
    m_1(s_1) &= 1 \\
    m_1(s_2) &= -1
\end{align*}

We now introduce the following assumption:

\begin{theorem}[Classicality]
\label{assump:class}
For every state $s_k$ and measurement $m_n$, there exists a function $f$ such that $f(n,k)=m_n(s_k)$.
\end{theorem}

We can think of this $f$ as a universal prediction machine for $\mathcal{S}$: given the index of a state and a measurement, it spits out the result the measurement will produce. For our coin example, this function is given by Table~\ref{tab:coin}:

 \begin{table}[h]
 \centering
  \caption{Measurement outcomes for a coin.} 
  
  \begin{tabular}{c | c c }\label{tab:coin}
 $f(n,k)$ & $s_1$   & $s_2$    \\ \hline
 $m_1$    &   1     &   -1     
 
  \end{tabular}
 \end{table}

Consequently, $f(1,1)=1$ (in state $s_1$, the coin shows heads), and $f(1,2)=-1$ (in state $s_2$, the coin shows tails). 

For the general case, with $i,j\in\mathbb{N}$, we obtain Table~\ref{tab:diag1}.

 \begin{table}[h]
 \centering
  \caption{Tabulation of the function $f(n,k)$ for a general system, together with an illustration of the diagonalization technique.} 
  
  \begin{tabular}{c | c c c c c c c c}\label{tab:diag1}
 $f(n,k)$ & $s_1$   & $s_2$    & $s_3$   & $s_4$   & $s_5$   &$\ldots$& $s_g$ &$\ldots$\\ \hline
 $m_1$  &   (1)   &    -1    &    1    &    1    &    1    & $\ldots$&     1    &$\ldots$\\
 $m_2$  &    1    &   (-1)   &    1    &    -1   &    -1   &         &    -1     &\\
 $m_3$  &    -1   &    1     &   (-1)  &    -1   &    -1   &         &     1     &\\
 $m_4$  &    1    &    -1    &    -1   &   (1)   &    1    &         &     1     &\\
 $m_5$  &    -1   &    -1    &    -1   &    1    &   (1)   &         &    -1     &\\
                    \vdots             &$\vdots$&          &          &          &         &$\xddots$ & $\vdots$ &\\
 $m_g$  &   -1    &     1    &    1    &    -1   &    -1   &$\ldots$ &    (\Lightning)    &$\ldots$\\
                    \vdots             &$\vdots$&          &          &          &         &          & $\vdots$&  $\xddots$\\
  \end{tabular}
 \end{table}

We can now lead Assumption~\ref{assump:class} to a contradiction. To do so, we must first observe that we can construct new measurements by means of logical operations. For this, it is convenient to think of the values $1$ and $-1$ as representing `true' and `false', respectively. Then, we can consider $m_n(s_k)=1$ to mean that the proposition `$\mathcal{S}$ has property $n$ in state $k$' is true, and $m_n(s_k)=-1$ consequently that it is false. Each measurement thus tests whether a system in a given state has or fails to have a certain property. Since properties and measurements are thus in one-to-one correspondence, we will on occasion abuse notation and speak of the `property $m_n$'.

We can equivalently look at this in terms of subsets (or -regions) of the state space introduced in Fig.~\ref{pic:phase}. Each measurement essentially tests whether the system is in some region of that space. For instance, the region with $p$ smaller than $\sqrt{2m E_0}$ corresponds to the set of states with energy $E$ less than $E_0$; a measurement that yields $1$ for all states in that region (and $-1$ otherwise) then indicates the truth of the proposition `$\mathcal{S}$ has energy less than $E_0$'.

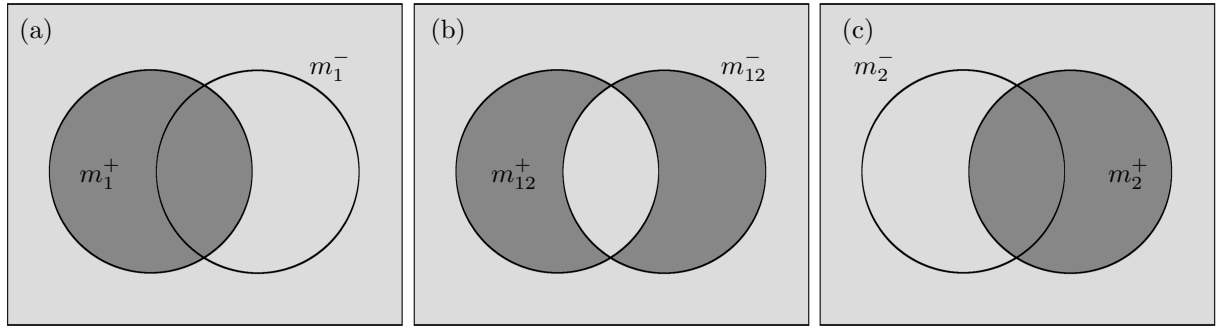
\begin{figure}[h] 
 \centering
 \begin{overpic}[width=\textwidth]{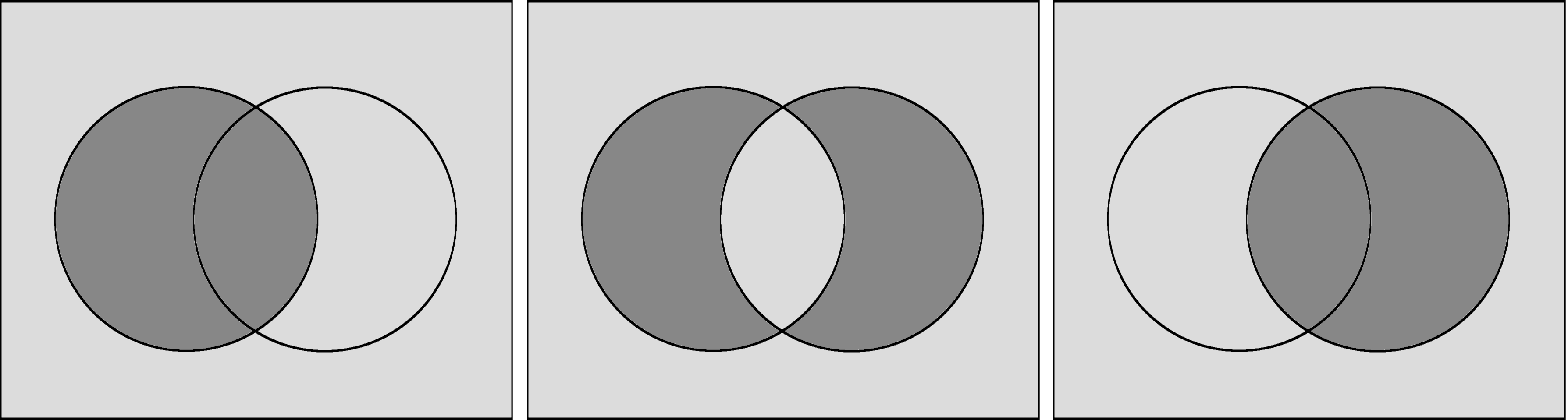}
  \put(1,24){(a)}
  \put(35,24){(b)}
  \put(69,24){(c)}
  \put(6,12){$m_1^+$}
  \put(40,12){$m_{12}^+$}
  \put(91,12){$m_2^+$}
  \put(25,21){$m_1^-$}
  \put(59,21){$m_{12}^-$}
  \put(70,21){$m_2^-$}
\end{overpic}
\caption{Property-calculus in phase space}
\label{pic:log}
\end{figure}

This enables us to construct a logical calculus for the properties of the system. From two measurements $m_1$ and $m_2$, we can, for instance, construct $m_{12}=m_1\oplus m_2$, where the operator $\oplus$ is taken to signify the logical xor: that is, $m_{12}=1$ if $m_{1}\neq m_{2}$, and $m_{12}=-1$ if $m_{1}= m_{2}$. For ease of notation, we indicate the property values by superscripts; see Fig.~\ref{pic:log}. 

Moreover, we can give an explicit measurement procedure for each property: simply measure momentum and position up to the precision necessary to localize the state within the respective subset.

But then, this means that we can construct the following measurement $m_g$: for each $m_i$, $m_g(s_i)$ is just the opposite of $m_i(s_i)$. That is, if $m_1(s_1)$ yields $1$, $m_g(s_1)$ yields $-1$; if $m_2(s_2)$ yields $-1$, then $m_g(s_2)$ yields $1$. The construction of this measurement is then shown in Table~\ref{tab:diag1}.

If we now hold fast to our assumption that $f(n,k)$ enumerates all possible measurement outcomes, then $m_g$ itself must correspond to some row of Table~\ref{tab:diag1}. However, it cannot correspond to the first row, as it differs from $m_1$ in the value associated to $s_1$; it cannot correspond to the second row, as it differs in the value associated to $s_2$; and so on, for any particular row of that table.

We might now suppose that, having infinitely many rows, we can just add the missing measurement. But, as is of course familiar, this move will not get us out of trouble: we can always just repeat the construction, finding a further measurement not on the list already. 

But this means that there exists some state $s_g$ and measurement $m_g$ such that the value of $m_g(s_g)$ cannot be predicted by $f$. Thus, our `universal prediction machine' cannot, in fact, exist; there are measurements such that their outcome for certain states cannot be predicted. They are, in other words, undecidable.

The above has the form of a \emph{diagonal argument}. Diagonalization was first introduced by Cantor in his famous proof of the existence of uncountable sets, and lies at the heart of G\"odel's (first) incompleteness theorem, the undecidability of the halting problem, and many others. The precise structure of such arguments in a category-theoretic setting was brought to the fore by Lawvere by means of a fixed-point theorem \cite{Law1969}. This can be directly adapted to the present setting, yielding a somewhat more general argument than the above; for details, see \cite{szangolies2018epistemic} and Appendix~\ref{app:lawvere}.

An intuitive way to understand this result is the following. Consider that we can define every measurement by listing the states that lie within the corresponding subregion of state space. Then, note that we can, correspondingly, define each state via measurements---say, listing all the measurements that yield a $+1$-outcome (all the properties the system possesses in that state). Thus, we can define a measurement in terms of a state defined in terms of that very measurement---yielding the paradoxical circularity characteristic of self-reference.

This has intriguing consequences. First of all, we cannot consistently assign to $s_g$ either a value of $1$ or $-1$ for $m_g$, as supposing it ought to be $1$ yields the conclusion that it must be $-1$, and vice versa. Thus, when faced with the question whether the system has property $m_g$, we find that we can neither affirm nor deny. This is, of course, just the situation Schr\"odinger's infamous and much-abused cat finds itself in: we can neither claim it is alive, nor that it is not. Thus, we may consider the system to be in a \emph{superposition} with respect to $m_g$. 

Suppose now we perform a measurement of $m_g$. Any possible outcome will be inconsistent with the system being in state $s_g$---since, as we had surmised, no outcome can consistently be associated with that state. Hence, after the measurement has yielded a result, it follows that the system can no longer be in the state $s_g$---that is, post measurement state change (`wave-function collapse') is a direct consequence of the preceding considerations.

It is important to note that quantum mechanics, itself, does not again fall prey to the same issues. There are two salient factors accounting for this: first, the proof depends on the possibility of `duplicating' the index $g$ to construct $m_g(s_g)$---which, physically, represents a cloning operation that is famously impossible in quantum mechanics \cite{WZ1982}. Second, we must be able to invert the value of a measurement---take the value $1$ to $-1$, and vice versa. That is, every possible value assigned to a property must be negated. 

But this likewise is impossible in quantum mechanics \cite{Sv1995}. Let us replace the classical outcomes with orthogonal quantum states $\ket{1}$ and $\ket{-1}$. Then, the operator
\begin{equation*}
    U_{NOT} = \ket{1}\bra{-1} + \ket{-1}\bra{1}
\end{equation*}
takes $\ket{1}$ to $\ket{-1}$, and vice versa. However, applied to the state $\frac{1}{\sqrt{2}}(\ket{1} + \ket{-1})$, we get
\begin{equation*}
    U_{NOT}\frac{1}{\sqrt{2}}(\ket{1} + \ket{-1}) = \frac{1}{\sqrt{2}}(\ket{1} + \ket{-1}).
\end{equation*}
Hence, the superposition yields a fixed point for $U_{NOT}$---thus evading the inconsistent assignment of Table~\ref{tab:diag1}. 

This is, of course, only the first whiff of quantum phenomena. The picture can be developed further. Complementarity, the impossibility to simultaneously assign definite values to certain properties, can be obtained by considering a form of the above argument in the context of sequences of measurements. Furthermore, the uncertainty principle emerges as a finite bound on the information available about a system---the number of simultaneously definite properties---by appealing to Chaitin's version of the incompleteness theorem \cite{Ch1992}. For details, see \cite{szangolies2018epistemic}. In the following, we will consider another paradigmatically quantum feature that has, so far, not been considered: entanglement.

For this, it will be useful to consider a simple `toy' system. Thus, take the extreme case of a system $\mathcal{S}$ such that only one of its properties is decidable---all save a single bit of information lies beyond the epistemic horizon. We are then in the situation of Fig.~\ref{pic:log}: one bit of information decides one of three possible mutually exclusive measurements on the system. Appreciating the parallel to the orthogonal measurements for a single qubit, we will name these three properties $x_\mathcal{S}$, $y_\mathcal{S}$, and $z_\mathcal{S}$.

\subsection{Entanglement}\label{subsec:entanglement}
So far, we have only considered single, individual systems. One might therefore ask what this framework entails once one investigates composite systems instead. Thus, take two systems, $\mathcal{A}$ and $\mathcal{B}$. To keep matters simple, we will confine our discussion here to `toy systems' of the kind introduced above---that is, systems described by a single definite property.

Consequently, $\mathcal{A}$ is described by either of $x_\mathcal{A}$, $y_\mathcal{A}$, or $z_\mathcal{A}$ having a definite value, while $\mathcal{B}$'s state is given by one out of $x_\mathcal{B}$, $y_\mathcal{B}$, and $z_\mathcal{B}$. A possible state of the compound system $\mathcal{A}\otimes\mathcal{B}$ would then be $(x_\mathcal{A}^+,x_\mathcal{B}^-)$, where we used the superscript notation to indicate property values. 

As we had surmised, however, we can use elementary Boolean logic to construct new properties. Thus, let us consider the property $x_\mathcal{AB} = x_\mathcal{A}\oplus x_\mathcal{B}$. This indicates a correlation between the two $x$-values: it signifies that one must be the opposite of the other. We can then e. g. give a complete description of the system as $(x_\mathcal{A}^+,x_\mathcal{AB}^+)$, signifying the state where $x_\mathcal{A} = 1$, and $x_\mathcal{B}$ must be the opposite, hence $-1$. 

This is, as yet, a completely classical situation. Picture the case of two colored cards, one red, and one green, in two envelopes: once you open one, you immediately know the color of the card within the other, even if the latter is located on Pluto. There is in particular nothing nonlocal about you having this knowledge.

However, consider now the state $(x_\mathcal{AB}^+,z_\mathcal{AB}^+)$. Here, the two bits of information we have available to describe the state are entirely taken up by the correlations: we know that the two $x$-values, as well as the two $z$-values, are opposed to one another; but we know nothing whatever about any individual $x$- or $z$-value!

This is precisely the situation of an entangled two-particle system (cf. \cite{zeilinger1999foundational}). Our $f(n,k)$, which, for this system, can only determine two properties, only provides values for $x_\mathcal{AB}$ and $z_\mathcal{AB}$, but leaves, e. g., $z_\mathcal{B}$ undecidable. However, once we have performed the requisite measurement, the considerations of the previous sections tell us that something remarkable must happen: whatever outcome is produced, one bit of information must now be taken up by the value of $z_\mathcal{B}$; but, due to the (anti-)correlation between $z$-values, this then immediately tells us the value of $z_\mathcal{A}$, as well! Furthermore, as all information available is now taken up by (e. g.) $(z_\mathcal{B}^+,z_\mathcal{AB}^+)$, it follows that nothing about the $x$-values can be known: the correlation there is destroyed.

\section{Does this Ring a Bell?}\label{sec:bell}
We now have the tools in hand to investigate one of the most famous expressions of quantum `weirdness': Bell's theorem \cite{bell1964einstein}, or the failure of `local realism'. We will start with a slightly different view on Bell inequalities \cite{cereceda2000local}.

Consider, to this end, again that there exists a function $f(n,k)$ providing values to all possible measurements. In particular, consider the above bipartite system and the properties $x_\mathcal{A}$, $z_\mathcal{A}$, $x_\mathcal{B}$ and $z_\mathcal{B}$. With respect to these properties, every state can be written as a four-tuple $(x_\mathcal{A},z_\mathcal{A},x_\mathcal{B},z_\mathcal{B})$, corresponding to a column in Table~\ref{tab:diag1}. That is, there are $16$ possible states, from $f(n,1) = (x_\mathcal{A}^+,z_\mathcal{A}^+,x_\mathcal{B}^+,z_\mathcal{B}^+)$ to $f(n,16) = (x_\mathcal{A}^-,z_\mathcal{A}^-,x_\mathcal{B}^-,z_\mathcal{B}^-)$, which we label $\lambda_i$. 

In any given experiment, each of these states may be present with a certain probability $P(\lambda_i) = p_i$. See Table~\ref{tab:prob} for an enumeration.

 \begin{table}[h]
 \centering
  \caption{States and probabilities for the bipartite syste $\mathcal{A}\otimes\mathcal{B}$}
  
  \begin{tabular}{c | c c c c | c }\label{tab:prob}
 State       & $x_\mathcal{A}$ & $z_\mathcal{A}$ & $x_\mathcal{B}$ & $z_\mathcal{B}$ & $P(\lambda_i)$    \\ \hline
 $\lambda_1$ &       1         &        1        &       1         &         1       & $p_1$             \\
 $\lambda_2$ &       1         &        1        &       1         &        -1       & $p_2$             \\
 $\lambda_3$ &       1         &        1        &      -1         &         1       & $p_3$             \\
 $\lambda_4$ &       1         &        1        &      -1         &        -1       & $p_4$             \\
 $\lambda_5$ &       1         &       -1        &       1         &         1       & $p_5$             \\
 $\lambda_6$ &       1         &       -1        &       1         &        -1       & $p_6$             \\
 $\lambda_7$ &       1         &       -1        &      -1         &         1       & $p_7$             \\
 $\lambda_8$ &       1         &       -1        &      -1         &        -1       & $p_8$             \\
 $\lambda_9$ &      -1         &        1        &       1         &         1       & $p_9$             \\
 $\lambda_{10}$&      -1         &        1        &       1         &        -1       & $p_{10}$            \\
 $\lambda_{11}$&      -1         &        1        &      -1         &         1       & $p_{11}$            \\
 $\lambda_{12}$&      -1         &        1        &      -1         &        -1       & $p_{12}$            \\
 $\lambda_{13}$&      -1         &       -1        &       1         &         1       & $p_{13}$            \\
 $\lambda_{14}$&      -1         &       -1        &       1         &        -1       & $p_{14}$            \\
 $\lambda_{15}$&      -1         &       -1        &      -1         &         1       & $p_{15}$            \\
 $\lambda_{16}$&      -1         &       -1        &      -1         &        -1       & $p_{16}$             
 
  \end{tabular}
 \end{table}

With this, we can compute probabilities for individual outcomes by marginalization---that is, summing over all probabilities for states that contain the desired outcome. Therefore, the probability to find $x_\mathcal{A} = 1$ is equal to $P(x_\mathcal{A}^+)=\sum_{i=1}^{8}p_i = p_1 + p_2 + \ldots + p_8$, as states $\lambda_1$ through $\lambda_8$ have $x_\mathcal{A}=1$. We can likewise compute probabilities for joint events: $P(x_\mathcal{A}^+,x_\mathcal{B}^-) = p_3 + p_4 + p_7 + p_8$. 

Finally, we can compute expectation values for such joint events: 
\begin{align*}
    \ex{x_\mathcal{A}x_\mathcal{B}} &= \sum_{r,s\in\{1,-1\}} rs P(x_\mathcal{A}^r,x_\mathcal{B}^s) \\
                                    &= P(x_\mathcal{A}^+,x_\mathcal{B}^+) + P(x_\mathcal{A}^-,x_\mathcal{B}^-) - P(x_\mathcal{A}^+,x_\mathcal{B}^-) - P(x_\mathcal{A}^-,x_\mathcal{B}^+) \\
                                    &= p_1 + p_2 - p_3 - p_4 + p_5 + p_6 - p_7 - p_8 - p_9 - p_{10} + p_{11} + p_{12} - p_{13} - p_{14} + p_{15} + p_{16}
\end{align*}

These expectation values carry information about the correlation between the two properties: if $\ex{x_\mathcal{A}x_\mathcal{B}} = 1$, only the $p_i$ with a positive sign are nonzero, and thus, $x_\mathcal{A} = x_\mathcal{B}$ for all states in the ensemble; for $\ex{x_\mathcal{A}x_\mathcal{B}} = -1$, we obtain $x_\mathcal{A} = -x_\mathcal{B}$. If $\ex{x_\mathcal{A}x_\mathcal{B}} = 0$, the value of $x_\mathcal{A}$ tells us nothing about $x_\mathcal{B}$, and vice versa.

With this, it is easy to compute the quantity
\begin{align*}
    \ex{C_{CHSH}} & = \ex{x_\mathcal{A}x_\mathcal{B}} + \ex{x_\mathcal{A}z_\mathcal{B}} + \ex{z_\mathcal{A}x_\mathcal{B}} - \ex{z_\mathcal{A}z_\mathcal{B}} \\
                  & = 2 - 4(p_3 + p_4 + p_6 + p_8 + p_9 + p_{11} + p_{13} + p_{14}) \\
                  & = 4(p_1 + p_2 + p_5 + p_7 + p_{10} + p_{12} + p_{15} + p_{16}) - 2.
\end{align*}

Since $\sum_i p_i \leq 1$, this immediately yields
\begin{equation*}
    -2 \leq \ex{C_{CHSH}} \leq 2.
\end{equation*}

This is, of course, nothing but the famous CHSH-Bell inequality \cite{clauser1969proposed}.

This should strike us as somewhat remarkable: only the assumption that there exists a $f(n,k)$ assigning values to all observables turns out to be enough to derive a bound on the above expression. Thus, Bell inequalities precisely delineate the set of theories for which there exists $f(n,k)$ such that it yields the values for all possible measurements. Contrariwise, Bell inequality violations certify that no such $f(n,k)$ for all values can exist---or at least, be probed by experiment.

The undecidability of these values then allows for the violation of this expression---as is, indeed, observed in quantum mechanics. Mathematically, the bounds on $\ex{C_{CHSH}}$ correspond to necessary conditions for the existence of a joint probability distribution (Table~\ref{tab:prob}); their violation means that no consistent assignment of probabilities to the $\lambda_i$ is possible.

This should not surprise us: we have already seen that, for instance, the event $(x_\mathcal{A}^+,z_\mathcal{A}^+)$ cannot occur---$f(n,k)$ does not assign simultaneous values to both elements. But if the above probability distribution were to exist, we could easily obtain
\begin{equation*}
    P(x_\mathcal{A}^+,z_\mathcal{A}^+) = p_1 + p_2 + p_3 + p_4.
\end{equation*}

But what could it mean to assign a probability to an impossible event?

It is more usual to attribute violations of Bell inequalities to the failure of either \emph{locality} or \emph{realism}. What does the above probability distribution have to do with either? 

`Realism' is ultimately simply the possibility of assigning values to all observables. If such an assignment is possible, each of the rows in Table~\ref{tab:prob} designates a valid state, and can be assigned a probability, leading to the above considerations.

But how is the failure of locality supposed to avoid this trouble? The resolution here is that we have implicitly assumed that we can fairly sample from the above probability distribution. However, if outcome probabilities on $\mathcal{B}$ were to change due to measurements on $\mathcal{A}$, then we could no longer carry the argument through. Hence, one usually makes an assumption that a measurement on one part of the system does not influence measurements carried out on the other; to make this assumption sensible, one ensures that both parts of the system are far away from one another, such that no influence, propagating at the speed of light, could travel between them. Should there then be any instantaneous influence despite these precautions, we speak of a failure of locality.

\section{EPistemic HoRizons: Incomplete Quantum Mechanics?}\label{sec:EPR}
It is sometimes proposed that Bell's theorem only hinges on the assumption of locality, and hence, its violation suffices to conclude that nature is nonlocal (e. g. \cite{maudlin2014bell}). The reasoning here is typically that `realism' is not a separate requirement that could fail on its own, but rather, is already established by the famous argument due to Einstein, Podolski, and Rosen (EPR) \cite{einstein1935can}. 

Let us take a lightning-quick review of the argument adapted to the present formalism. EPR take a system in the state $(x_\mathcal{AB}^+,z_\mathcal{AB}^+)$, and consider measurements on one of its parts (say $\mathcal{A}$). Upon measuring $x_\mathcal{A}$, we obtain the $x$-value for $\mathcal{A}$, and due to the correlation given by $x_\mathcal{AB}^+$, can immediately infer $x_\mathcal{B}$; likewise for $z$. However, the quantum formalism does not permit us to speak of simultaneous values for $x_\mathcal{B}$ and $z_\mathcal{B}$. But how, then, is $\mathcal{B}$ supposed to know to `produce' the right value in each case?

The EPR-argument hinges on a bit of \emph{counterfactual reasoning}: \emph{had} we measured $z_\mathcal{A}$ (instead of $x_\mathcal{A}$), we \emph{would have} been able to predict a definite value for $z_\mathcal{B}$ (instead of $x_\mathcal{B}$). Due to the absence of any disturbance on $\mathcal{B}$ due to our actions on $\mathcal{A}$ (locality), we then conclude that $\mathcal{B}$ cannot just spontaneously `decide' which value to produce, and hence, both $x_\mathcal{B}$ and $z_\mathcal{B}$ must have had a definite value---in EPR's parlance, an `element of reality'---associated to them all along. 

To illustrate this puzzle, Schr\"odinger introduced the analogy of the fatigued student \cite{schrodinger1935discussion}: quizzed in an oral examination, they will get the first answer right with certainty, after which, however, any further answer will be random. Even though we only get one correct answer out in any case, we still must conclude that the student knew the answer to every question, in order to produce this performance: \emph{had} we asked a different first question, then nevertheless the student \emph{would have} produced the right answer.

Applied to quantum mechanics, this would entail that the description of the correlated system $\mathcal{A}\otimes\mathcal{B}$ must be incomplete: $\mathcal{B}$ must, to give the right answer in each of these cases, `know' the correct values for $x_\mathcal{B}$ and $z_\mathcal{B}$ in advance, these answers simply being hidden to the quantum formalism.

If this is correct, then nonlocality is our only out in the case of Bell's theorem: there are simultaneous values for all observables---$f(n,k)$ does not tell the whole story---and measuring one part of a system must influence the value distribution of the distant part.

One way to attempt to defuse the force of EPR's argument is to deny that the sort of counterfactual inference that allows us to reason about what would have happened had our measurement choice been different is valid, at least in a quantum context. However, without further substantiation regarding why that should be the case, simply denying the validity of a certain form of argument to avoid an unwelcome conclusion hardly seems fair.

While I do not presume to settle this controversy once and for all, I believe the present framework offers a fresh perspective on the matter. For consider what happens in each of the two cases. The initial state $(x_\mathcal{AB}^+,z_\mathcal{AB}^+)$ becomes, say, $(z_\mathcal{A}^+,z_\mathcal{AB}^+)$, respectively $(x_\mathcal{A}^+,x_\mathcal{AB}^+)$. The only change is thus in the properties of the local system $\mathcal{A}$, about which we have gained new information. 

This allows us then to \emph{infer} the value of the distant system. However, we may hold that this is something different than that value spontaneously becoming definite---after all, this value is not given by any $f(n,k)$. We could thus associate `elements of reality' to the values of $f(n,k)$ exclusively. Our conclusions about the distant values would then have the status of inferences about the truth value of the G\"odel sentence: We can \emph{infer} that `I am not provable (in a given axiomatic system)' is true, since it is, in fact, not provable (in that system); however, the system itself will not be able to establish this truth (on pain of contradiction).

Such a state is one in which we have the following two items of knowledge: `the $x$/$z$-value of $\mathcal{A}$ is 1' and `the $x$/$z$-value of $\mathcal{B}$ is opposite that of $\mathcal{A}$'. This differs from a state like $(x_\mathcal{A}^+,x_\mathcal{B}^-)$ in a subtle, but crucial, way. In that state, our knowledge is given by `the $x$-value of $\mathcal{A}$ is 1' and `the $x$-value of $\mathcal{B}$ is -1'. The difference emerges if we imagine varying the first of each set of propositions---that is, engage in counterfactual reasoning. In case of a state like $(x_\mathcal{A}^+,x_\mathcal{B}^-)$, we can say that \emph{had} we obtained a value of $1$ for the $z$-value instead, we could still validly speak of the $x$-value of $\mathcal{B}$ being $-1$. 

That is not the case for the state $(x_\mathcal{A}^+,x_\mathcal{AB}^+)$: varying the first proposition, but leaving the second constant, would lead us to a state in which we have \emph{no information} about the $x$-value of $\mathcal{B}$. Consequently, the two states differ in the counterfactuals they support: the state $(x_\mathcal{A}^+,x_\mathcal{B}^-)$ allows us to say that, \emph{had} the first value been different, the second \emph{would have been the same} (absent any disturbance), leading to e. g. $(z_\mathcal{A}^+,x_\mathcal{B}^-)$. However, in the state $(x_\mathcal{A}^+,x_\mathcal{AB}^+)$, as soon as we imagine exchanging $x_\mathcal{A}$, we loose any ability to make determinations of $x_\mathcal{B}$, as this value is specified only contingently on that of $x_\mathcal{A}$. In a state like `$(z_\mathcal{A}^+, x_\mathcal{AB}^+)$', $x_\mathcal{B}$ would simply not have any determinate value at all. 

An alternative way to think about the situation is by introducing the notion of a \emph{conditional event}. A conditional event is, for instance, an observable that only takes a value conditionally on the value of another. Thus, we can write the state $(x_\mathcal{A}^+,x_\mathcal{AB}^+)$ equivalently as $(x_\mathcal{A}^+,x_\mathcal{B}^-|x_\mathcal{A}^+)$, where the `$|$'-notation denotes the conditioning: the value of $x_\mathcal{B}$ is $-1$ \emph{given that} the value of $x_\mathcal{A}$ is $+1$. This is a rewriting of the information contained in $x_\mathcal{AB}^+$ that more clearly emphasizes the result of $\mathcal{A}$ observing a given value on their ability to predict the value of $\mathcal{B}$'s measurement. 

We should then not think about the state $(x_\mathcal{A}^+,x_\mathcal{B}^-|x_\mathcal{A}^+)$ as containing the value of $\mathcal{B}$'s measurement; rather, the two bits of information it contains jointly entail $\mathcal{B}$'s value. This is a salient difference: consider, for example, the case of a one-time pad: one bit of the encrypted message plus one bit of the key may entail one bit of the clear text, in the same way that $\mathcal{A}$'s measurement result, plus knowledge of the correlation, entails $\mathcal{B}$'s outcome. But that does not mean that the state, as such, must contain information about $\mathcal{B}$'s value if $\mathcal{A}$'s value were different, anymore than the value of the key alone must contain information about the clear text.

Thus, only given that one has actually measured $x_\mathcal{A}$ is reasoning about the value of $x_\mathcal{B}$ possible. In this sense, the present framework gives a natural meaning to Bohr's somewhat opaque `influence on the precise conditions which define the possible types of prediction which regard the subsequent behaviour of the system' \cite{bohr1935can}. We naturally imagine it to be possible to change one thing, while keeping something else equal; but in this case, the `one thing' (the definite value of $x_\mathcal{A}$) is part of the antecedent conditions for making determinations about that `something else' (the value of $x_\mathcal{B}$). Moreover, trying to simultaneously appeal to both $(x_\mathcal{A}^+,x_\mathcal{B}^-|x_\mathcal{A}^+)$ and $(z_\mathcal{A}^+,z_\mathcal{B}^-|z_\mathcal{A}^+)$ (for instance) amounts to exceeding the information bound on the system as given by $f(n,k)$; thus, the illegitimate nature of the counterfactual argument in this case is seen to be rooted in the more fundamental informational limit. The epistemic horizon puts a limit to the information accessible about any given system, and each attempt to access more courts inconsistency.

The EPR argument, then, essentially trades on a conflation of $(x_\mathcal{A}^+,x_\mathcal{B}^-|x_\mathcal{A}^+)$ with $(x_\mathcal{A}^+,x_\mathcal{B}^-)$. Only the latter state supports the reasoning that leads us to conclude that the distant particle must have `known' the value of both $x_\mathcal{B}$ and $z_\mathcal{B}$ all along.

\section{Hardy's Paradox}\label{sec:hard}

The tools developed above can be fruitfully applied to other supposed `paradoxes' in the quantum world. Consider, to this end, the Hardy state \cite{hardy1992quantum, hardy1993nonlocality} of two entangled qubits, which is in the $z_\mathcal{A}z_\mathcal{B}$-basis

\begin{equation}\label{one}
    \ket{\psi_H}=\frac{1}{\sqrt{3}}\left(\ket{z_\mathcal{A}^+z_\mathcal{B}^+} + \ket{z_\mathcal{A}^+z_\mathcal{B}^-} + \ket{z_\mathcal{A}^-z_\mathcal{B}^+} \right).
\end{equation}

Here, a state such as $\ket{z_\mathcal{A}^+z_\mathcal{B}^+}$ means that $\mathcal{A}$ and $\mathcal{B}$ would obtain the values $z_\mathcal{A}^+$ resp. $z_\mathcal{B}^+$ upon performing the requisite measurements. 

Hardy's paradox now consists in pointing out that elementary reasoning suffices to demonstrate that two parties, $\mathcal{A}$ and $\mathcal{B}$, measuring each qubit in the $x$-basis $\{\ket{x^+},\ket{x^-}\}=\{\frac{1}{\sqrt{2}}(\ket{z^+} + \ket{z^-}),\frac{1}{\sqrt{2}}(\ket{z^+} - \ket{z^-})\}$, can never both see the state $\ket{x^-}$ (i. e. obtain the outcomes $x_\mathcal{A}^-$ and $x_\mathcal{B}^-$). Yet, in fact, this happens with a probability $p_H = \frac{1}{12}$.

This can be seen by writing $\ket{\psi_H}$ in the $x_\mathcal{A}x_\mathcal{B}$-basis. This yields:

\begin{equation}\label{two}
    \ket{\psi_H}= \frac{3}{\sqrt{12}}\ket{x_\mathcal{A}^+x_\mathcal{B}^+} + \frac{1}{\sqrt{12}}\ket{x_\mathcal{A}^+x_\mathcal{B}^-} + \frac{1}{\sqrt{12}}\ket{x_\mathcal{A}^-x_\mathcal{B}^+} - \frac{1}{\sqrt{12}}\ket{x_\mathcal{A}^-x_\mathcal{B}^-}
\end{equation}

According to the Born rule, measurements performed on this state yield the $\ket{x_\mathcal{A}^-x_\mathcal{B}^-}$-outcome with probability $p_H=\frac{1}{12}$. 

It is useful, here, to look at the chain of reasoning used to arrive at the above conclusion in greater detail (cf. \cite{dourdent2020quantum}). To start with, in the above notation, for the Hardy state in the $z_\mathcal{A}z_\mathcal{B}$-basis, the state contains the information that `if $\mathcal{B}$ obtains the outcome $z_\mathcal{B}^-$, then $\mathcal{A}$ obtains the outcome $z_\mathcal{A}^+$'; once $\mathcal{B}$ then obtains that outcome, we have the information content $(z_\mathcal{B}^-,z_\mathcal{A}^+|z_\mathcal{B}^-)$. 

As before, the notation `$z_\mathcal{A}^+|z_\mathcal{B}^-$' expresses the conditional nature of $\mathcal{A}$'s value; only \emph{given that} $\mathcal{B}$ obtained the value $z_\mathcal{B}^-$ can we consistently talk about $\mathcal{A}$'s observed value. 

Now, the state in the $x_\mathcal{A}z_\mathcal{B}$-basis is:

\begin{equation}\label{three}
    \ket{\psi_H} = \sqrt{\frac{2}{3}}\ket{x_\mathcal{A}^+z_\mathcal{B}^+} + \frac{1}{\sqrt{6}}\ket{x_\mathcal{A}^+z_\mathcal{B}^-} + \frac{1}{\sqrt{6}}\ket{x_\mathcal{A}^-z_\mathcal{B}^-}
\end{equation}

From this, we see that, if $\mathcal{A}$ measures $x_\mathcal{A}=-1$, $\mathcal{B}$ must obtain $z_\mathcal{B}=-1$, that is, the information within the state afterwards is $(x_\mathcal{A}^-,z_\mathcal{B}^-|x_\mathcal{A}^-)$.

Finally, in the $z_\mathcal{A}x_\mathcal{B}$-basis, the state is:

\begin{equation}\label{four}
    \ket{\psi_H} = \sqrt{\frac{2}{3}}\ket{z_\mathcal{A}^+x_\mathcal{B}^+} + \frac{1}{\sqrt{6}}\ket{z_\mathcal{A}^-x_\mathcal{B}^+} + \frac{1}{\sqrt{6}}\ket{z_\mathcal{A}^-x_\mathcal{B}^-}
\end{equation}

If $\mathcal{A}$ thus obtains $z_\mathcal{A}=+1$, $\mathcal{B}$ must obtain $x_\mathcal{B}=+1$, and the information content afterwards is $(z_\mathcal{A}^+,x_\mathcal{B}^+|z_\mathcal{A}^+)$.

This now suffices to establish the contradiction. Suppose we were to reason as follows:
\begin{enumerate}[(i)]
    \item\label{i} If $\mathcal{A}$ obtains $x_\mathcal{A}^-$, we can conclude that $\mathcal{B}$ must obtain $z_\mathcal{B}^-$, due to \ref{three}.
    \item\label{ii} Thus, suppose $\mathcal{B}$ then in fact obtains $z_\mathcal{B}^-$. With \ref{one}, we can then conclude that $\mathcal{A}$, had she measured in the $z_\mathcal{A}$-basis, would obtain $z_\mathcal{A}^+$.
    \item\label{iii} However, if $\mathcal{A}$ obtains $z_\mathcal{A}^+$, then \ref{four} tells us that $\mathcal{B}$ must obtain $x_\mathcal{B}^+$.
    \item\label{iv} (From \ref{i} - \ref{iii}) Putting these together, we surmise that if $\mathcal{A}$ obtains $x_\mathcal{A}^-$, $\mathcal{B}$ must obtain $x_\mathcal{B}^+$, and consequently, the outcome $\ket{x_\mathcal{A}^-x_\mathcal{B}^-}$ can never occur.
    \item\label{v} Yet, by \ref{two}, $\ket{x_\mathcal{A}^-x_\mathcal{B}^-}$ occurs with probability $p=\frac{1}{12}$. \Lightning
\end{enumerate}

To see what goes wrong here, let us go back, for a moment, to the discussion of the EPR paradox. There, we surmised that the information within a state $(x_\mathcal{A}^+,x_\mathcal{B}^-|x_\mathcal{A}^+)$ crucially differs from that in a state like $(x_\mathcal{A}^+,x_\mathcal{B}^-)$ in that the latter, but not the former, supports counterfactual inferences. That is, if we have the information about both systems individually, we can imagine varying the value of one system independently; but if the information about one system is only specified conditionally on that of the other, then counterfactual reasoning becomes nonsensical.

The EPR argument would successfully establish the incompleteness of quantum mechanics if, when $\mathcal{A}$ measures in the $x$-basis, we had the state $(x_\mathcal{A}^+,x_\mathcal{B}^-)$, and likewise, for $z$-measurement, the state $(z_\mathcal{A}^+,z_\mathcal{B}^-)$. For then, we could say that if $\mathcal{A}$ \emph{had measured} in a basis different from the one in which she actually did measure in any given experiment, $\mathcal{B}$'s particle nevertheless must have been prepared to produce a fitting answer. These two states could hence be termed counterfactually consistent, and we can appeal to both in a single argument.

However, that is not the case for states of the form $(x_\mathcal{A}^+,x_\mathcal{B}^-|x_\mathcal{A}^+)$. Here, $\mathcal{B}$'s value is only specified conditional on $\mathcal{A}$'s; thus, we cannot consistently imagine varying only $\mathcal{A}$'s value, as it forms part of the determining conditions of $\mathcal{B}$'s value. The states $(x_\mathcal{A}^+,x_\mathcal{B}^-|x_\mathcal{A}^+)$ and its counterpart $(z_\mathcal{A}^+,z_\mathcal{B}^-|z_\mathcal{A}^+)$ are thus not counterfactually consistent, and cannot be used in a single argument.

But the same is then true for $(x_\mathcal{A}^-,z_\mathcal{B}^+|x_\mathcal{A}^-)$ and $(z_\mathcal{A}^+,x_\mathcal{B}^+|z_\mathcal{A}^+)$. Both apply only in the contexts in which $\mathcal{A}$ did, in fact, make the $x$- respectively $z$-basis measurement. Since $\mathcal{A}$ cannot in fact make both measurements, propositions (\ref{i}) and (\ref{iii}) cannot simultaneously be appealed to: their combination would exceed the amount of information consistently obtainable about the system.

Consequently, the `paradox' in the above argument is of the same nature as that due to EPR, and similarly tells us that, in a quantum world, we must be careful which propositions about a system are simultaneously definite, and thus, can be used to underwrite counterfactual arguments.

\section{The Frauchiger-Renner Argument}\label{sec:fra}

Recently, an intriguing new argument has been presented by Daniela Frauchiger and Renato Renner \cite{frauchiger2018quantum}. They aim to show that ``quantum theory cannot consistently describe the use of itself", and use an ingenious thought experiment to support their claim. The paper has already received much commentary, which points both to the high impact and controversial nature of their result, as well as to the lack of consensus regarding its interpretation. 

The Frauchiger-Renner argument can be read as a `Wigner's Friendification' \cite{aaronson2018think} of Hardy's paradox. In a famous Gedankenexperiment \cite{wigner1995remarks}, Wigner ($\mathcal{W}$) asks us to imagine a hermetically sealed laboratory containing a scientist (the eponymous `Friend' $\mathcal{F}$) carrying out a Schrödinger's cat-type experiment. At some point, $\mathcal{F}$ will have made some definite observation of the cat's well-being. Yet, $\mathcal{W}$, having no knowledge of $\mathcal{F}$'s result (although he may have knowledge that $\mathcal{F}$ has observed \emph{some} definite result), must, applying the usual rules of quantum mechanics, describe the entire laboratory system as being in a state of superposition. Indeed, in theory, he could perform an interference experiment on the entire laboratory that would confirm his description.

But this poses a problem: $\mathcal{F}$, we should expect, has made a definite observation, yet $\mathcal{W}$'s description and experimental results are incompatible with any given definite state of the laboratory system. 

The `Wigner's Friend'-scenario is essentially a `Wigner's Friendification' of the EPR-argument: the latter features two entangled systems, while the former makes one of these systems a conscious observer, and adds another observer (a `meta-observer', \cite{dourdent2020quantum}) which carries out a measurement on the total system in an orthogonal basis. $\mathcal{W}$, we imagine, knows that the system is either in the state (cat alive, friend sees cat alive) or (cat dead, friend sees cat dead)---since both are incompatible with interference, we conclude, there must be some contradiction. Perhaps $\mathcal{F}$'s observation collapses the wave function, and thus, standard quantum rules no longer obtain once conscious observation is involved.

However, crucially, according to the above discussion, $\mathcal{W}$ in fact only knows that the system is in the state (cat alive, friend sees cat alive$|$cat alive) or (cat dead, friend sees cat dead$|$cat dead). And these, we had surmised, cannot be simultaneously appealed to consistently. Hence, the conclusion of a contradiction does not, in fact, obtain.

Frauchiger and Renner now essentially formulate a Wigner's-Friendified version of the Hardy paradox: consider two observers, $\mathcal{A}$'s friend $\mathcal{F_A}$ and $\mathcal{B}$'s friend $\mathcal{F_B}$, which share an entangled two-qubit system, and perform $z$-basis measurements on their respective qubits. In the state

\begin{equation*}
    \ket{\psi_H}=\frac{1}{\sqrt{3}}\left(\ket{z_\mathcal{A}^+z_\mathcal{B}^+} + \ket{z_\mathcal{A}^+z_\mathcal{B}^-} + \ket{z_\mathcal{A}^-z_\mathcal{B}^+} \right),
\end{equation*}

we now consider, e. g., $z_\mathcal{A}^+$ to be the `belief state' of $\mathcal{A}$'s friend $\mathcal{F_A}$ after performing a $z$-measurement and obtaining the outcome $+1$---analogous to $\mathcal{F}$'s state after observing the cat. $\mathcal{A}$ and $\mathcal{B}$ then carry out their measurements on the entire laboratories containing their respective friends in the basis $\{\ket{x^+},\ket{x^-}\}=\{\frac{1}{\sqrt{2}}(\ket{z^+} + \ket{z^-}),\frac{1}{\sqrt{2}}(\ket{z^+} - \ket{z^-})\}$, as before. However, this is now to be interpreted as a measurement testing for the superposed states of the entire laboratories, containing their respective friends, here labeled by their respective `belief states'. 

As before, simple application of the Born rule immediately tells us that both $\mathcal{A}$ and $\mathcal{B}$ may observe the $-1$-outcome with probability $\frac{1}{12}$. We can now again apply the reasoning of Hardy's paradox to obtain the apparent contradiction. However, in this version, the argument has an added wrinkle: we are not merely thinking about results $\mathcal{A}$ (say) \emph{would have} obtained, \emph{had} she made the appropriate measurements, but about measurements actually performed by $\mathcal{F}_\mathcal{A}$. Does this change matters?

From \ref{three}, we find that, if $\mathcal{A}$ obtains $-1$, $\mathcal{F_B}$ must obtain $-1$, likewise. But then, if $\mathcal{F_B}$ obtains $-1$, \ref{one} tells us that $\mathcal{F_A}$ must obtain the $+1$-outcome. Finally, \ref{four} tells us that given that $\mathcal{F_A}$ sees $+1$, $\mathcal{B}$ must obtain the $+1$-outcome.

In summary: having obtained the value $-1$ in her measurement, $\mathcal{A}$ knows that $\mathcal{F_B}$ knows that $\mathcal{F_A}$ knows that $\mathcal{B}$ must obtain the value $+1$, and thus, knows herself that $\mathcal{B}$ must obtain the value $+1$; yet, with probability $\frac{1}{12}$, both $\mathcal{A}$ and $\mathcal{B}$ obtain the outcome $-1$. 

The point of the Wigner's-Friendification is then the following: we are now not considering different measurements that $\mathcal{A}$ \emph{could have} performed (but didn't), but rather, measurements as actually performed by distinct observers, who presumably have each obtained definite measurement results. It is then tempting to think of these as `facts in the world', available for classical---that is, Boolean---logical reasoning. 

\begin{figure}[h] 
 \centering
 \begin{overpic}[width=0.8\textwidth]{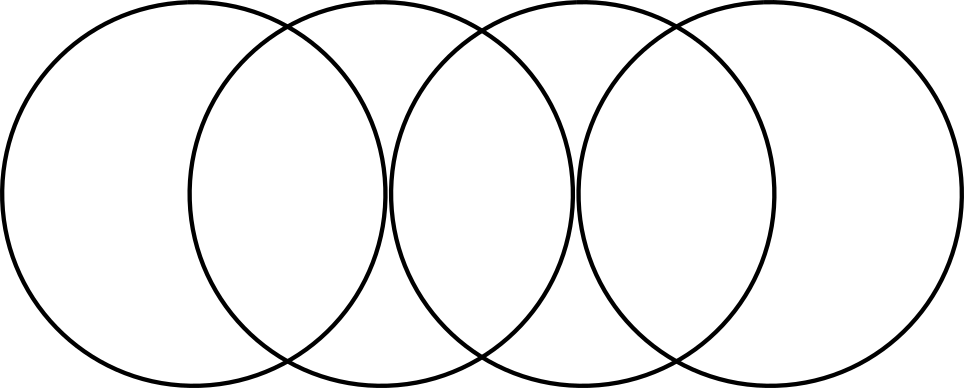}
  \put(19,36){$\mathcal{A}$}
  \put(38,36){$\mathcal{F_B}$}
  \put(58,36){$\mathcal{F_A}$}
  \put(79,36){$\mathcal{B}$}
  \put(5,19){$x_\mathcal{A}=-1$}
  \put(25,19){$z_\mathcal{B}=-1$}
  \put(45,19){$z_\mathcal{A}=+1$}
  \put(65,19){$x_\mathcal{B}=+1$}
\end{overpic}
\caption{Epistemic horizons of the observers in the Frauchiger-Renner argument}
\label{pic:hori}
\end{figure}
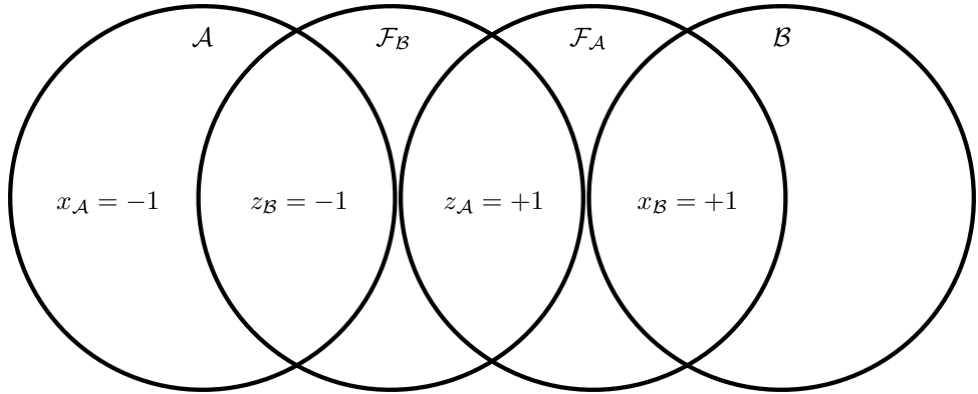

But there still is no unified logical framework encompassing both $(x_\mathcal{A}^-,z_\mathcal{B}^-|x_\mathcal{A}^-)$ and $(z_\mathcal{A}^+,x_\mathcal{B}^+|z_\mathcal{A}^+)$---appealing to both simultaneously amounts to exceeding the information limitation about the system as given by $f(n,k)$. $\mathcal{F_B}$'s observation of $-1$ can be contained within $\mathcal{A}$'s epistemic horizon, but $\mathcal{F_B}$'s determination of $\mathcal{F_A}$'s observation cannot also be: as shown in Fig.~\ref{pic:hori}, each individual epistemic horizon contains at most two bits of information. The Frauchiger-Renner argument then exceeds that limit by trying to unite the different, overlapping horizons into one---an impossibility already highlighted by the impossibility of finding a joint probability distribution over all observables in a Bell experiment. Hence, the `telescoping' of knowledge necessary for $\mathcal{A}$'s conclusion that $\mathcal{B}$ can never see $-1$ if she sees $-1$ cannot be performed: $\mathcal{A}$'s attempt to peek behind her epistemic horizon fails.

Frauchiger and Renner codify this `telescoping' in their assumption $C$, which says that ``a theory $T$ must [...] allow any agent $A$ to promote the conclusions drawn by another agent $A'$ to his own conclusions" \cite{frauchiger2018quantum}. This assumption, then, fails to be satisfied, if the preceding framework is apt. However, this is not an instance of quantum theory failing to ``consistently describe the use of itself"; rather, quantum theory, as already established by EPR and Hardy-type arguments, restricts which propositions can be consistently combined, without exceeding the bound on the maximal information that can be contained within a system.

One important lesson of the Frauchiger-Renner argument then is that it is not the counterfactual reasoning, per se, that is problematic in quantum mechanics, but rather, exceeding the informational limitation given by the undecidable values of $f(n,k)$. In the EPR and Hardy-arguments, this limitation is exceeded via counterfactually appealing to values that would have been obtained, had different measurements been carried out; but even if, as in the case of the FR argument, these measurements are actually performed, the bound on the maximum information available for any given system prohibits appealing to them within a single argumentative context.

\section{Conclusion}\label{sec:conc}
We have considered the application of self-referential arguments to physical systems, and found that many paradigmatically quantum phenomena seem to gain a natural explanation from this perspective. This idea is not entirely new: John Wheeler himself proposed the undecidable propositions of mathematical logic as a candidate for a `quantum principle', from which to derive the phenomenology of quantum mechanics \cite{whe1974}---a proposal which, as legend has it, got him thrown out of G\"odel's office \cite{ber1991}. For a brief review of these efforts, see \cite{szangolies2018epistemic} and references therein.

What this program, if successful, shows is that there is a common thread behind mathematical undecidability and physical unknowability---that, in other words, the epistemic horizons the pure mathematician and the experimental physicist find delimiting their perspectives are not separated, but instead, spring from a common source. 

In an intriguing sense, the incompleteness of mathematics may then come to the rescue of physics, allowing it in turn to yield a complete picture: the incompleteness the EPR-argument seeks to establish is averted by the horizon that bars counterfactual reasoning about unperformed experiments---which, hence, famously `have no results' \cite{peres1978unperformed}. It is as if Schr\"odinger's student does not know the answer to any questions, as such, but knows each answer only relative to that question being asked.

This motivates a proposal of \emph{relative realism}: assign `elements of reality' only where $f(n,k)$ yields a definite value. In this way, we get as close to the classical ideal of local realism as is possible in a quantum world. The resolution of the EPR, Hardy, and Frauchiger-Renninger paradoxes is then to deny the EPR notion of `elements of reality': according to their definition, an element of reality is associated with every value that can be predicted with certainty. But in a state such as $(x_\mathcal{A}^+,x_\mathcal{B}^-|x_\mathcal{A}^+)$, we can predict $x_\mathcal{B}^-$ with certainty, but no element of reality is associated to it; rather, it is the conditional value $x_\mathcal{B}^-|x_\mathcal{A}^+$ that is definite in this sense, which does not allow us to make any determination of $x_\mathcal{B}$ in the absence of a definite value for $x_\mathcal{A}$, and which cannot stand for $x_\mathcal{B}^-$ in chains of inferences.

We may try, combining indirectly-obtained information from different contexts in ever more ingenious ways, to look beyond our epistemic horizon; but the Old One's secrets, it seems, are not so easily discerned.

\newpage
\printbibliography[heading=bibintoc]

\newpage
\begin{appendices}
\section{The Lawvere Fixed-Point Argument}\label{app:lawvere}

We will explicitly construct a measurement $m_g(s_k)$, that is, a function $m_g:\Sigma_\mathcal{S} \to\{1,-1\}$, where $\Sigma_\mathcal{S}$ denotes the state space of $\mathcal{S}$, such that it differs from $f(n,k)$ for at least one $s_k$. 

Suppose that there exists a function $f(n,k):\mathbb{N}\times\mathbb{N}\to\{1,-1\}$ such that it is equal to the outcome of the $n$th measurement for the $k$th state. Furthermore, we introduce the arbitrary map $\alpha:\{1,-1\}\to\{1,-1\}$, and the map $\Delta:\mathbb{N}\to\mathbb{N}\times\mathbb{N}$ that takes $n \in \mathbb{N}$ to the tuple $(n,n)\in\mathbb{N}\times\mathbb{N}$. With these, we construct $g$ as the map that makes the following diagram commute:

\[
\begin{tikzcd}[column sep=large, row sep=huge]
\mathbb{N}\times \mathbb{N} \arrow{r}{f} & \{1,-1\} \arrow{d}{\alpha} \\
\mathbb{N} \arrow{u}{\Delta} \arrow{r}{g} & \{1,-1\} \\
\end{tikzcd}
\]

The map $g$ constructed in this way then yields sequentially values for a certain measurement, $m_g$, if performed on states of $\mathcal{S}$, i.e. $g(k)=m_g(s_k)$. If $f$ yields the value of every measurement applied to every state, then there must be some $n$ such that $g(k)=f(n,k)$ for all states $s_k$. Choose now $k=n$ and evaluate $g(n)$:

\begin{align*}
f(n,n)	& =g(n) \\
		& =\alpha(f(n,n))
\end{align*}

The first equality is simply our stipulation that $g$ should encode some measurement, and that $f(n,n)$ yields the outcome of the $n$th measurement on the $n$th state. The above then shows that the map $\alpha$ must have a fixed point at $f(n,n)$ for the construction to be consistent.

However, we are free in our choice of $\alpha$, and consequently, may choose the negation $\neg(1)=-1$, $\neg(-1)=1$. But this clearly has no fixed point, and we obtain the contradiction
\begin{equation*}
    f(n,n) = \neg f(n,n)\hspace{0.5cm} \text{\Lightning}
\end{equation*}

But then, this means that no $f$ reproducing every measurement outcome can exist.
\end{appendices}
\end{document}